\documentstyle[epsfig,a4]{article}

\begin{document}

\renewcommand{\thesection}{\arabic{section}.}
\renewcommand{\figurename}{\small Fig.}
\renewcommand{\theequation}{\arabic{section}.\arabic{equation}}

\newcommand{\eqreset}{\setcounter{equation}{0}}

\newcommand{\gtrsim}{
\,\raisebox{0.35ex}{$>$}
\hspace{-1.7ex}\raisebox{-0.65ex}{$\sim$}\,
}

\newcommand{\lesssim}{
\,\raisebox{0.35ex}{$<$}
\hspace{-1.7ex}\raisebox{-0.65ex}{$\sim$}\,
}

\newcommand{\onehalf}{\mbox{\scriptsize 
\raisebox{1.5mm}{1}\hspace{-2.7mm}
\raisebox{0.5mm}{$-$}\hspace{-2.8mm}
\raisebox{-0.9mm}{2}\hspace{-0.7mm}
\normalsize }}

\newcommand{\onefourth}{\mbox{\scriptsize 
\raisebox{1.5mm}{1}\hspace{-2.7mm}
\raisebox{0.5mm}{$-$}\hspace{-2.8mm}
\raisebox{-0.9mm}{4}\hspace{-0.7mm}
\normalsize }}

\bibliographystyle{prsty}

\begin{flushleft}

{\small {\em submitted to}\\
 J. Phys. A: Math. Gen. }
\vspace{2cm}

{
\Large\bf
Ordering in magnetic films with surface anisotropy
}      

\vspace{0.1cm}

\end{flushleft}

\begin{flushright}
\parbox[t]{11cm}{

D A Garanin$^\dagger$
\vspace{0.1cm}

\small

Max-Planck-Institut f\"ur Physik komplexer Systeme, N\"othnitzer Strasse 38,
D-01187 Dresden, Germany\\ 
\vspace{0.2cm}

Received 8 October 1998

\vspace{0.4cm}

{\bf Abstract.} \hspace{1mm}
Effects of the surface exchange anisotropy on 
ordering of ferromagnetic films are studied for the exactly solvable classical
spin-vector model with $D\to\infty$ components.
For small surface anisotropy $\eta'_s\ll 1$ (defined relative to the exchange
interaction), the shift of $T_c$ in a film
consisting of $N\gg 1$ layers behaves as $T_c^{\rm bulk} - T_c(N) \propto (1/N)
\ln(1/\eta'_s)$ in three dimensions.
The finite-size-scaling limit $T_c^{\rm bulk} - T_c(N) \propto 1/( \eta'^{1/2}N^2)$,
which is realized for the model with a bulk anisotropy $\eta'\ll 1$ in the range
$N\eta'^{1/2} \gtrsim 1$, never appears for the model with the pure
surface anisotropy. 
Here for $N\exp(-1/\eta'_s) \gtrsim 1$ in three dimensions, film orders at a
temperature above $T_c^{\rm bulk}$ (the surface phase transition). 
In the semi-infinite geometry, the surface phase transition occurs for
whatever small values of  $\eta'_s$  
(i.e., the special phase transition corresponds to $T_c^{\rm bulk}$) 
in dimensions three and lower. 
} 
\vspace{0.8cm}

\end{flushright}

\renewcommand{\thefootnote}{\fnsymbol{footnote}}
\renewcommand{\footnoterule}{\rule{0cm}{0cm}}
\footnotetext[2]{http://mpipks-dresden.mpg.de/$\sim$garanin/; e-mail:
garanin@mpipks-dresden.mpg.de, \\ garanin@physnet.uni-hamburg.de, garanin@t-online.de}

\section{Introduction}
\eqreset

Reduction of Curie temperatures $T_c$ of ferromagnetic films consisting of
$N\gg 1$ layers with respect to the bulk value is usually
represented in the form
%
\begin{equation}\label{TcShNbig}
[T_c^{\rm bulk} - T_c(N)]/T_c^{\rm bulk} \cong A/N^\lambda. 
\end{equation}
For the exponent $\lambda$ the finite-size scaling theory
\cite{fisbar72,bar83ptcp} yields
$\lambda=1/\nu_b$, where  $\nu_b$ is critical index for the bulk correlation length.
The above result has been derived with the Ising model, or the field model
with one-component order parameter, in mind.
For weakly anisotropic Heisenberg model and, in general, for models with several
spin components, the nearly Goldstone modes can drastically change the
character of ordering in magnetic films.
In particular, in the dimensionality range $d\leq 3$ in the isotropic limit at low
temperatures, the film behaves as a
system of dimensionality $d'=d-1 \leq 2$  and cannot order because of long-wavelength fluctuations.
This means that the amplitude $A$ in Eq.\ (\ref{TcShNbig}) should diverge in
the isotropic limit.
Moreover, even the functional form of Eq.\ (\ref{TcShNbig}) should change to
reflect explicitly the $d'$ dimensional nature of a nearly isotropic film.
For the model with the uniaxial exchange anisotropy (longitudinal spin
components coupled by $J$ and transverse components coupled by $\eta J$ with $\eta \leq
1$, so that $\eta' \equiv 1-\eta$ measures the anisotropy) it was shown in
Refs.\ \cite{gar96jpal,gar98sffb} that Eq.\ (\ref{TcShNbig}) is only valid for rather thick films,
$N\kappa_c \gtrsim 1$, where 
$\kappa_c \equiv 1/\xi_{c\perp} =\sqrt{2d(1/\eta-1)}$ is the inverse
transverse correlation length at the bulk critical point, which goes to zero
in the isotropic limit.
Here in three dimensions  for the classical spin-vector model
with $D\to\infty$ components one has $\lambda=2$ and $A \sim 1/\kappa_c$.
In the range $N\kappa_c \lesssim 1$ a $d'$ dimensional behavior is realized, which
is characterized by $\lambda=1$ and $A \sim \ln[1/(\kappa_cN)]$ in three
dimensions, $d=3$.
For extremely small anisotropies, the film orders at such low temperatures
that spins along the direction perpendicular to the surface are strongly
correlated with each other and they can be considered as single composite spins.
Thus the film is mapped on $d'$-dimensional monolayer with the exchange
interaction $NJ$, which yields \cite{gar98sffb}
%
\begin{equation}\label{TcAnLo}
T_c(d,J,\eta',N) \cong T_c(d',NJ,d\eta'/d',1).
\end{equation}
Although the results of Refs.\ \cite{gar96jpal,gar98sffb} have
been obtained for the infinite-component classical vector model,
 the qualitative features of this solution should be shared by
the more realistic Heisenberg model, $D=3$.
In particular, formula (\ref{TcAnLo}) is model independent and valid for all
$D\geq 2$.

The purpose of this paper is to study ordering in magnetic films with a
{\em surface} anisotropy.
The latter arises, typically, due to the violation of the symmetry of a crystal field
acting on the magnetic ions at the surface.
Although this anisotropy has a single-site form, we will consider here the
anisotropy of the exchange interactions between the surfaces spins, instead.
This leads to the same qualitative results and allows one to use the formalism
developed for the exchange-anisotropy models in
Refs.\ \cite{gar96jpa,gar96jpal,gar98pre,gar98sffb}.
One can expect that the surface anisotropy stabilizes ordering in films with
$d\leq 3$ and $N \gg 1$ weaker than the bulk one.
If the surface anisotropy is very small, then $T_c \ll T_c^{\rm bulk}$, and at
such low temperature, its influence should redistribute over $N$ layers,
so that its effective value be $\eta'_{\rm eff} \sim \eta'_s/N$.
The latter should result in a more pronounced suppression of $T_c$ in magnetic films. 
This can be immediately seen in the case of extremely small surface
anisotropy, where the analogue of Eq.\ (\ref{TcAnLo}) reads
%
\begin{equation}\label{TcAnSurfLo}
T_c(d,J,\eta'_s,N) \cong T_c(d',NJ,\eta'_s/N,1).
\end{equation}

A specific feature of the model with pure surface anisotropy is the absence of the finite
length scale, such as the transverse correlation length $\xi_{c,\perp} \equiv
1/\kappa$, at criticality.
As a result, the system is always in the range $N\kappa \ll 1$ and there is no
crossover to the finite-size-scaling regime of Eq.\ (\ref{TcShNbig}).   
As a result, for a small surface anisotropy, the corresponding analytical solution for the
Curie temperature of the film holds in a much wider range of $N$.

If surface anisotropy exceeds a critical value $\eta_{s,c}(N)$, the Curie
temperature of the film exceeds the bulk Curie temperature:
$T_c > T_c^{\rm bulk}$.
The possibility of this effect, which is absent in the mean field
approximation (MFA), can be seen from the following simple arguments.
The isotropic large-$D$ model orders at $T_c^{\rm bulk} = J_0/(DW_d)$, where
$J_0$ is the zero Fourier component of the exchange interaction and 
$W_d\equiv P_d(1)$ [see Eq. (\ref{thetacbulk})] is
the Watson integral containing the information on the lattice dimensionality
and structure.
On the other hand, the Curie temperature of the monolayer with the (surface)
anisotropy of the extreme Ising type, $\eta_s =0$ (i.e.,  $\eta'_s=1$), is    
$T_c(1) = (d'/d)J_0/D$.
For the simple cubic lattice one has $W_3=1.51639$, so that the Curie
temperature of the anisotropic monolayer slightly exceeds the isotropic bulk Curie
temperature.
That is, the lack of interacting neighbours at the surface can be compensated for
by a stronger suppression of $T_c^{\rm bulk}$ due to long-wavelength
fluctuations making contribution to $W_d$.
It is clear that the bilayer has a substantially higher value of $T_c$ than
the monolayer, and that in dimensions lower than 3 the bulk Curie temperature
is suppressed even stronger.
For the continuous-dimension model introduced in
Ref.\ \cite{gar98pre}, one has $W_{3.0}=1.719324$ and $W_{2.5}=2.527059$.
In two dimensions and below,, $W_d$ diverges and thus $T_c^{\rm bulk}$ goes to zero in the
isotropic limit.
On the other hand, the theory predicts a finite-temperature surface phase transition
 for any nonzero values of the surface anisotropy $\eta'_s$.
Thus, for $d\leq 2$  the surface anisotropy  is the only source of ordering.
This situation is realized only in the limit $D\to\infty$, however.
Since in two dimensions the surface is one dimensional,
ordering at the surface should be destroyed by thermal fluctuations of the
longitudinal spin components  for any finite $D$.       
In fact, surface anisotropy plays a major role already for $d\leq 3$.
We will see below that $\eta_{s,c}(N)$ goes to zero in the limit $N\to\infty$
in this dimensionality range.
Thus in the semi-infinite geometry a surface phase
transition above $T_c^{\rm bulk}$ occurs for whatever small value of
the surface anisotropy, i.e.,  the bulk Curie temperature is the temperature
of the special phase transition as well! 

The main part of this paper is organized as follows.
In Sec.\ \ref{sec_basic} the closed system of equations describing the $D\to
\infty$ component spin-vector model in the symmetric phase is written down.
In Sec.\ \ref{sec_AnisSmall} the analytical calculation of the correction to
$T_c$ in films with a weak surface anisotropy is presented.
In Sec.\ \ref{sec_AnisLarge} the surface phase transition is considered.
The results of numerical calculations are at appropriate places in sections
above.
In Sec.\ \ref{sec_Disc} the results obtained are summarized, and possibilities
of finding similar regimes in more realistic models are discussed.

\section{Basic equations and their solution}
\label{sec_basic}
\eqreset

The Hamiltonian of the anisotropic classical $D$-component spin-vector model can be written in the form
%
\begin{equation}\label{dham}
{\cal H} = 
- \frac{1}{2}\sum_{ij}J_{ij}
\left(
m_{zi}m_{zj}  
+ \eta_{ij} \sum_{\alpha=2}^D m_{\alpha i} m_{\alpha j}
\right) , 
\qquad |{\bf m}_i|=1,
\end{equation}
where dimensionless anisotropy factors satisfy $\eta_{ij} \leq 1$.
This model was introduced, in the isotropic form, by Stanley, who showed that
its partition function in the spatially homogeneous case in the limit
$D\to\infty$ \cite{sta68pr} coincides with that of the spherical model
\cite{berkac52}.
There are, however, a number of essential differences between the exactly
solvable limit $D\to\infty$  of Eq.\ (\ref{dham}) and the spherical model.
In particular, there is only one correlation function (CF)  in the spherical model,
and thus this model cannot incorporate anisotropy.
In the $D\to\infty$ model, there are longitudinal and transverse CFs which
differ below $T_c$, even in the spatially homogeneous isotropic case \cite{gar97zpb}.

The system of equations describing the spatially inhomogeneous $D\to\infty$ model both above and below
$T_c$ was obtained in Ref.\ \cite{gar96jpa}.
At or above $T_c$ in zero field, the magnetization  $\langle{\bf m}_i\rangle$
is zero and the model is
described by the closed system of equations for the correlation functions of transverse
($\alpha\geq 2$) spin components, $s_{ij}\equiv D\langle m_{\alpha i}m_{\alpha j}\rangle$, 
and the spatially varying gap parameter, $G_i$.
(The definition of $G_i$ can be found in Ref.\ \cite{gar96jpa}; here it is
nonessential.)
In the film geometry,  it is 
convenient to use the Fourier representation in $d'=d-1$ translationally invariant dimensions parallel to the surface and the site representation in the $d$th dimension. 
The equations can be easily generalized for the anisotropy factors taking the
values $\eta_{nn}$ within the $n$th layer and $\eta_{n,n\pm 1}$ for the
interaction between the $n$th and  $(n\pm 1)$th layers.
For the model  with nearest-neighbor (nn) interactions,
the equation for the Fourier-transformed CF $\sigma_{nn'}({\bf q})$ 
then takes the form of a system of  second-order finite-difference
equations in the set of layers $n=1,\,2,...,N$:
%
\begin{equation}\label{CFfd}
2b_n\eta_{nn}\sigma_{nn'} - \eta_{n,n+1}\sigma_{n+1,n'} - \eta_{n,n-1}\sigma_{n-1,n'}  =
2d\theta\delta_{nn'} ,
\end{equation}
where $b_n$ is given by
%
\begin{equation}\label{bn}
b_n = d/(\eta_{nn} G_n) - d'\lambda'_{\bf q} , 
\end{equation}
$\lambda'_{\bf q}$ for the $d$-dimensional hypercubic lattice reads 
%
\begin{equation}\label{lampr}
\lambda'_{\bf q} = \frac{1}{d'}\sum_{i=1}^{d'} \cos(q_i) ,
\end{equation}
and the lattice spacing has been set to unity.
In Eq.\ (\ref{CFfd}), $\theta$ is the reduced temperature defined by
%
\begin{equation}\label{thetadef}
\theta \equiv \frac{ T }{ T_c^{\rm MFA}(\infty) } , 
\qquad T_c^{\rm MFA}(\infty) = \frac{ J_0 }{ D },
\end{equation}
where for hypercubic lattices $J_0=2dJ$.
The quantities $\sigma_{0,n'}$ and $\sigma_{N+1,n'}$ in the nonexisting
layers, which enter equations
(\ref{CFfd}) at the film boundaries $n=1$ and $n=N$, are set to
%
\begin{equation}\label{bcond}
\sigma_{0,n'} = \sigma_{N+1,n'} = 0    
\end{equation}
as free  boundary conditions.
The autocorrelation functions in each of $N$ layers, $s_{nn}$, satisfy the
set of constraint equations
%
\begin{equation}\label{sconstrfd}
 s_{nn} \equiv \int\!\!\!\frac{d^{d'}{\bf q}}{(2\pi)^{d'}} 
\sigma_{nn}({\bf q}) = 1 , 
\end{equation}
which are the consequence of the spin rigidity, $|{\bf m}_i|=1$.
A straightforward algorithm for numerical solving the equations above
is to compute, for a given set of $G_n$,  all $\sigma_{nn}$ from the system of
linear equations (\ref{CFfd}) and then insert the results in
Eq.\ (\ref{sconstrfd}) to obtain, after the integration over the Brillouin zone, a set of nonlinear
equations for $G_n$.

The first step of the routine described above can be conveniently done with the help of the
continued-fraction formalism which is described in detail in
Refs.\ \cite{gar98pre,gar98sffb}.
For a particular type of the model with surface and
bulk anisotropies, which is defined by
%
\begin{equation}\label{DefSurfAnis}
\eta_{11}=\eta_{NN} = \eta_s \leq 1, \qquad \eta_{nn}=\eta_{n,n\pm 1} = \eta
\leq 1
\qquad (nn \neq 11,NN),
\end{equation}
and which will be studied below,  it is convenient to rewrite equations (\ref{CFfd}) in the form
%
\begin{equation}\label{CFfdMo}
2\tilde b_n\sigma_{nn'} - \sigma_{n+1,n'} - \sigma_{n-1,n'}  =
(2d\theta/\eta) \delta_{nn'} ,
\end{equation}
where $\tilde b_n = (\eta_s/\eta) b_n$ for $n=1,N$ and $\tilde b_n = b_n$ otherwise.
Explicitly,
%
\begin{equation}\label{bnMo}
\tilde b_n = d/(\eta G_n) - d'\lambda'_{\bf q} + 
( 1 -  \eta_s / \eta )d'\lambda'_{\bf q} (\delta_{n,1} + \delta_{nN} ). 
\end{equation}

An alternative way to find $\sigma_{nn'}$, which is more appropriate for the
analytical treatment,  is to represent equations
(\ref{CFfd}) in the matrix form 
%
\begin{equation}\label{CFfdMatr}
\hat B \hat \sigma = {\rm diag}(2d\theta/\eta_{nn}),
\qquad B_{nn} = 2b_n, \qquad B_{n,n\pm 1} = -\eta_{n,n\pm 1}/\eta_{nn},
\end{equation}
so that the solution for $\sigma_{nn'}$ is given by
$\sigma_{nn'} = (2d\theta/\eta_{n'n'}) B_{nn'}^{-1}$.
Since the diagonal part of the matrix $\hat B$, which depends on the wave
vector ${\bf q}$, is proportional to the unity matrix, the eigenvalues and
eigenvectors of $\hat B$ can be defined as 
%
\begin{equation}\label{EigenB}
\hat B \hat U_\rho = [ \mu_\rho + 2d'(1-\lambda'_{\bf q})] \hat U_\rho,
\qquad \rho = 1,2, \ldots, N,
\end{equation}
the eigenvectors $ \hat U_\rho$ being independent of ${\bf q}$.
It should be noted that matrix $\hat B$ is nonsymmetric, 
$ B_{n,n\pm 1} = -\eta_{n,n\pm 1}/\eta_{nn} \neq  
B_{n\pm 1,n} = -\eta_{n,n\pm 1}/\eta_{n\pm1,n\pm1}$, if anisotropy factors
$\eta_{nn}$ change from one layer to the other.
In this case its left eigenvectors $\hat W_\rho^T$  differ from its right eigenvectors
$\hat U_\rho$.
The Green function $\sigma_{nn'}$ can be expanded over the set of eigenvectors
of the problem as follows
%
\begin{equation}\label{signn'mat}
\sigma_{nn'}({\bf q}) = 
\frac{2 d\theta }{ \eta_{n'n'} } \sum_{\rho=1}^N \frac{ U_{n\rho} W_{\rho n'}^{T} }
{ \mu_\rho + 2d'(1-\lambda'_{\bf q}) }.
\end{equation}
Here  matrix $\hat U$ is composed of the right eigenvectors $\hat U_\rho$ as columns
and $\hat W^T$  is composed of the left eigenvectors $\hat W_\rho^T$ as raws.
The right and left eigenvectors satisfy the biorthogonality condition 
$\sum_n W_{n\rho} U_{n\rho'} = \delta_{\rho\rho'}$.
In general,  matrix  $\hat U$ in nonunitary: $\hat U^{-1} = \hat W^T \neq \hat
U^T$.
Integration in Eq.\ (\ref{sconstrfd}) can be performed analytically with the result
%
\begin{equation}\label{SConstrMat}
s_{nn} = \frac{2 d\theta }{ \eta_{nn} } \sum_{\rho=1}^N 
\frac{ U_{n\rho} W_{\rho n}^{T} }{ 2d' +\mu_\rho } 
P_{d'}\left(\frac{ 2d' }{ 2d' + \mu_\rho } \right) 
= 1 , 
\end{equation}
where $P_{d'}(X)$ is the lattice Green function for the layer, which is
defined similarly to the lattice Green function $P(X)\equiv P_d(X)$ below.
Using this method with tabulated values of $P_{d'}(X)$ can save
computer time, in comparison to the continued-fraction method.
On the other hand, the continued-fraction method is fast enough and already
implemented, so that it will be used here. 
The diagonalization formalism above will be used for analytically solving the
problem in the next two sections.

After the set of $G_n$  for a given temperature has been determined, one can compute the longitudinal CF
$\sigma_{nn'}^{zz}({\bf q})$ from Eqs.\ (\ref{CFfd}) and  (\ref{bn}), where all
anisotropy factors $\eta_{nn'}$ are replaced by 1.
The Curie temperature of the film $\theta_c$ can now be found from the equation
%
\begin{equation}\label{TcEq}
[\sigma_{nn}^{zz}({\bf q}=0)]^{-1} = 0.
\end{equation}
In a usual situation, the above condition should be used in the middle of
the film, $n \sim N/2$, because for large $N$ the critical divergence of the spin CF at the surface
is suppressed  \cite{bramoo77prl,bramoo77jpa,gar98pre}.
If ordering of the film is driven by the surface, it is more convenient to use
Eq.\ (\ref{TcEq}) for $n=1$.
This equation has, in general, $N$ roots, as we will see below.
One should choose the maximal root for $\theta_c$, all other roots are
unphysical.
Below  $\theta_c$, the spontaneous magnetization appears, and the
very form of the equations change.

One can also represent $\sigma_{nn'}^{zz}$  in the form of
Eq.\ (\ref{signn'mat}) with $\eta_{n'n'}\Rightarrow 1$, where eigenvalues $\lambda_\rho^z$ and eigenvectors
components $U_{n\rho}^z$ correspond to the problem with the matrix  $\hat B^z$.
The latter is defined by Eq.\ (\ref{CFfdMatr}),  where anisotropy factors $\eta_{nn'}$ are replaced by 1.
Since $\hat B^z$ is a real symmetric matrix, $(\hat B^z)^T=\hat B^z$, matrix   $\hat U^z$ is unitary: $ (\hat U^z)^{-1} =
( \hat U^z)^T$, i.e., $U_{\rho n}^{z,-1} = U_{n\rho}^z$.
The eigenvalue problem corresponding to the longitudinal CF  can be written in
the form of a discrete Schr\"odinger equation for a particle with mass $m=1/2$:
%
\begin{equation}\label{SchrEq}
- \psi_{n-1} + 2 \psi_n - \psi_{n+1} + V_n \psi_n = E \psi_n,
\qquad V_n =  2d(1/G_n -1),
\end{equation}
as in  quantum tight-binding models.
This form is useful for the interpretation of the results; the eigenvectors and
eigenvalues of Eq.\ (\ref{SchrEq}) are more compact forms of the quantities
introduced above:
%
\begin{equation}\label{sigzzmat}
\sigma_{nn'}^{zz}({\bf q}) = 2d\theta \sum_{\rho=1}^{N}  \frac{
\psi_{n\rho}\psi_{n'\rho} }{ E_\rho + q^2 }, \qquad q \ll 1,
\end{equation}
where $E_\rho\equiv \mu_\rho^z$ and $ \psi_{n\rho}\equiv U_{n\rho}^z $.
The condition for the Curie temperature of the film has the form 
%
\begin{equation}\label{TcEqMat}
E_1(\theta_c)=0,
\end{equation}
where $E_1$ is the lowest of the eigenvalues $E_\rho$.
The $N-1$ solutions corresponding to $E_\rho=0$, $\rho\geq 2$, are unphysical. 
It should be noted that for the transverse correlation function the problem cannot, in
general, be interpreted quantum mechanically, since the matrix $\hat B$ may
be non-Hermitean, as is the case for the model with surface anisotropy.
The eigenvalues of the transverse problem, $\mu_\rho$, exceed the longitudinal
eigenvalues $E_\rho$; in the Ising limit $\eta \ll 1$ one has $\mu_\rho \propto
1/\eta$, whereas $E_\rho$ become independent of $\eta$.   

One should note that the longitudinal CF is in our formalism only a ``slave'' quantity, it does not
affect the basic equations of the model and is not subject to a constraint
condition similar to Eq.\ (\ref{sconstrfd}).
The physical reason for that is irrelevance of fluctuations of the {\em single}
longitudinal component in comparison to those of $D-1$ transverse ones in the
limit $D\to\infty$.

In the spatially homogeneous bulk sample one has $G_n=G$ and
$\eta_{nn}=\eta_{n,n\pm 1} =\eta$, and the transverse CF can be easily found
 \cite{gar98pre,gar98sffb}.
The resulting equation for the gap parameter has the form
%
\begin{equation}\label{GbulkEq}
\theta G P(\eta G) = 1,
\end{equation}
where 
%
\begin{equation}\label{Px}
P(X) \equiv \int\!\!\!\frac{d^d{\bf k}}{(2\pi)^d}\frac{1}{1-X\lambda_{\bf k}} 
\end{equation}
is the lattice Green function.
The quantity $\lambda_{\bf k} \equiv J_{\bf k}/J_0$ for the nearest-neighbour
interaction  is given by Eq.\ (\ref{lampr}) with $d'\Rightarrow d$ and ${\bf q} \Rightarrow \bf k$.
The solution $G$ of Eq.\ (\ref{GbulkEq}) increases with lowering temperature
$\theta$; at $G=1$ the gap in the longitudinal CF closes, longitudinal
susceptibility diverges, and the phase transition occurs.
This defines the bulk transition temperature  \cite{garlut84d}
%
\begin{equation}\label{thetacbulk}
\theta_c^{\rm bulk} = 1/P(\eta),
\end{equation}
that generalizes the well known result for the spherical model 
$\theta_c=1/P(1)$ \cite{berkac52}.
The lattice Green function $P(X)$ satisfies $P(0)=1$ and has a singularity at $X\to 1$, the form of
which in different dimensions can be found in Ref.\ \cite{gar98pre}.
For $d\leq 2$, the Watson integral $W\equiv P(1)$ goes to infinity; thus
formula (\ref{thetacbulk}) yields nonzero values of the Curie temperature only
for the anisotropic model, $\eta<1$.
It should be noted that in the anisotropic case the critical indices of the
model coincide with the mean-field ones due to the suppression of the
singularity of $P(\eta G)$ for $G\to 1$.
Below $\theta_c$, the spontaneous magnetization appears, and $G$ sticks to 1.
 
In Eq.\ (\ref{Px}) one has $\lambda_{\bf k} \cong 1-k^2/(2d)$ in the
long-wavelength limit.
Thus the inverse transverse correlation length $\kappa$ following from Eq.\ (\ref{Px})
is defined by
%
\begin{equation}\label{kapdef}
\kappa^2 \equiv 2d[1/(\eta G) -1] .
\end{equation}
Its critical-point value $\kappa_c  \equiv \sqrt{ 2d[1/\eta -1]}$ measures the
bulk anisotropy and
varies between 0 for the isotropic model and $\infty$ for the classical Ising model.
The inverse longitudinal correlation length $\kappa_z$ is determined by 
$\kappa_z^2 \equiv 2d[1/G -1]$ and it diverges at the critical point.
In contrast to finite-$D$ theories, where the longitudinal correlation length 
$\xi_{cz}\equiv 1/\kappa_z$ plays the major role in the scaling, here in the limit $D\to\infty$ it becomes
only a slave variable, whereas all the physical quantities, except the
longitudinal CF,   are scaled with the
transverse correlation length  $\xi_{c,\perp} \equiv 1/\kappa$  \cite{gar98pre,gar98sffb}.

\begin{figure}[t]
\unitlength1cm
\begin{picture}(15,7.5)
\centerline{\epsfig{file=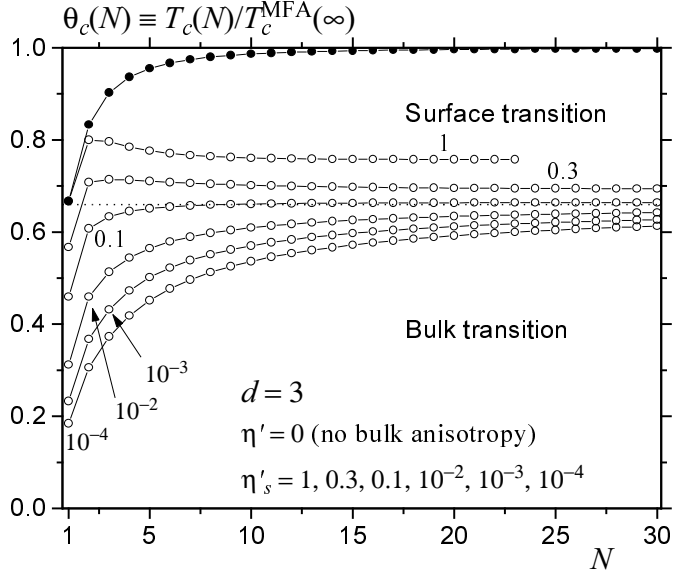,angle=-90,width=10cm}}
\end{picture}
%
\caption{ \label{s3tclin}
Curie temperatures of the $N$-layer simple-cubic-lattice film for different
values of surface anisotropy.
Horizontal dotted line is the bulk value of $T_c$.
Solid circles are the values of $T_c$ for the model with the bulk anisotropy
$\eta' =1$ (classical Ising model).  
}
\end{figure}

Numerical solution of the problem with the method described in this
section above yields the results for $\theta_c(N)$ of the three-dimensional film
with a simple cubic structure, which are shown in Fig.\ \ref{s3tclin}.
One can see that for small transverse anisotropies, $\theta_c(N)$ approaches
its bulk limit much slowlyer than the solution for the model with bulk
anisotropy $\eta=0$ (classical Ising model), which is shown by solid circles.
Since in the latter case transverse spin components are switched off, the
result coincides with that of the mean field approximation \cite{woldewhalpal71}
%
\begin{equation}\label{TcMFA}
\theta_c(N) = 1 - \frac 1 d \left( 1 - \cos{\frac{\pi}{N+1}} \right) 
\cong  1 - \frac{ 1 }{ 2d } \left( \frac \pi N \right)^2  \quad (N\gg 1).
\end{equation}

As can be seen from Fig.\ \ref{s3tclin}, the suppression of the Curie
temperature in films with a weak surface anisotropy may be quite pronounced,
especially in comparison with the mean-field result shown by solid circles.
Large $T_c$ shifts in films have been observed in many experiments (see, e.g.,
Ref.\ \cite{libab92}).
For larger values of $\eta'_s$,  the film orders above $T_c^{\rm bulk}$.
This is an indication that ordering at the surface occures first
 and thus determines the Curie temperature of the film.
The decrease of $\theta_c(N)$ with $N$ in this region can be easily explained.
For $N\to\infty$ both surfaces order independently  at some
$\theta_c(\infty)$.
For finite $N$, the surfaces interact with each other across the film and thus
help each other to order.
The interaction between surfaces, and thus the corresponding increase of
$\theta_c$, should decay exponentially with the film
thicknes $N$, the characteristic length being the bulk correlation length. 
One can see that for $\eta'_s=1$ surfaces order at a temperature substantially
higher than $\theta_c^{\rm bulk}$, where the  bulk correlation length is
rather short.
With lowering $\eta'_s$, the bulk correlation length at $\theta_c$ increases,
and the effect of the interaction of surfaces becomes more and more
pronounced.
The mechanism described above will be considered in more detail in
Sec.\ \ref{sec_AnisLarge}

\begin{figure}[t]
\unitlength1cm
\begin{picture}(15,7.5)
\centerline{\epsfig{file=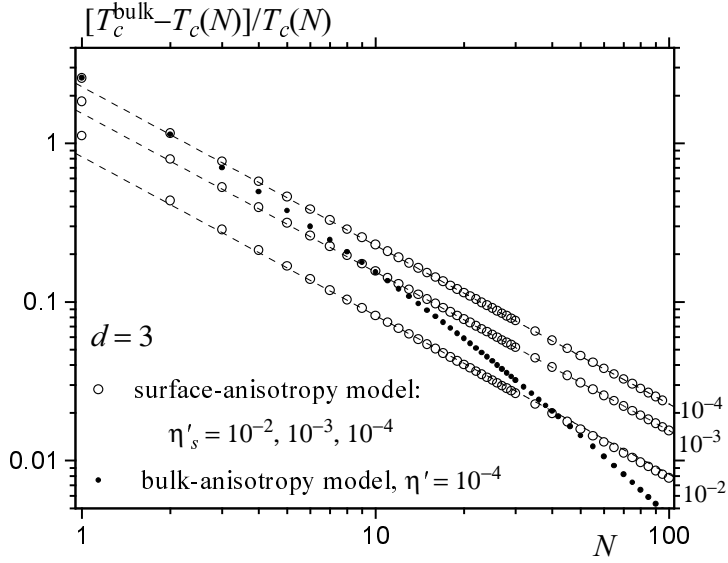,angle=-90,width=10cm}}
\end{picture}
%
\caption{ \label{s3tclog}
Curie-temperature shifts in simple-cubic-lattice films for small
values of surface anisotropy.
Dashed lines represent Eq.\ (\protect\ref{TcShiftSC}) with $c_3=1.36$.
}
\end{figure}

The Curie temperatures of films consisting of one and two layers can be
calculated analytically since there is no inhomogeneity of the gap parameter
$G_n$.
For the monolayer the result can be obtained by a straightforward
renormalization of Eq.\ (\ref{thetacbulk}) and has the form 
$\theta_c^{-1} = [d/(d-1)] P_{d'}(\eta)$
(there is no difference between the models with bulk and surface anisotropies).
For the bilayer, the surface-anisotropy model orders, evidently, at lower temperatures
than the bulk-anisotropy one.
For the latter, the expression for $\theta_c$ can be found in
Ref.\ \cite{gar98sffb}.
For the surface-anisotropy model, the result has the form
%
\begin{equation}\label{TcBilayerSA}
\theta_c^{-1} = \frac 12 \left[
 \frac{ d }{ d' } P_{d'}(\eta_s) + P_{d'}\left(\frac{ d' }{ d }\eta_s\right)
\right].
\end{equation}
For $\eta_s=0$ one has $P=1$, and this formula yields $\theta_c =
2(d-1)/(2d-1)$, which becomes 4/5 for $d=3$ (see Fig.\ \ref{s3tclin}).
For comparison, for the model with the bulk anisotropy $\eta=1$, the mean-field
formula (\ref{TcMFA}) yields  $\theta_c = (2d-1)/(2d)$ for $N=2$.
This becomes 5/6 for $d=3$ (see Fig.\ \ref{s3tclin}).
An interesting feature of the solution for the surface-anisotropy model is
that the Curie temperature of the bilayer becomes independent of the lattice
structure in the Ising limit $\eta_s=0$.
The result obtained above depends on the lattice dimensionality $d$ only and,
e.g.,  it is the same for the simple cubic model ($d=3$) and the
three-dimensional continuous-dimension model ($d=3.0$) \cite{gar98pre}.
In the Ising limit, the lattice structure comes into play for trilayers and thicker films,
where the inhomogeneity of the gap parameter $G_n$ becomes essential.

\begin{figure}[t]
\unitlength1cm
\begin{picture}(15,7.5)
\centerline{\epsfig{file=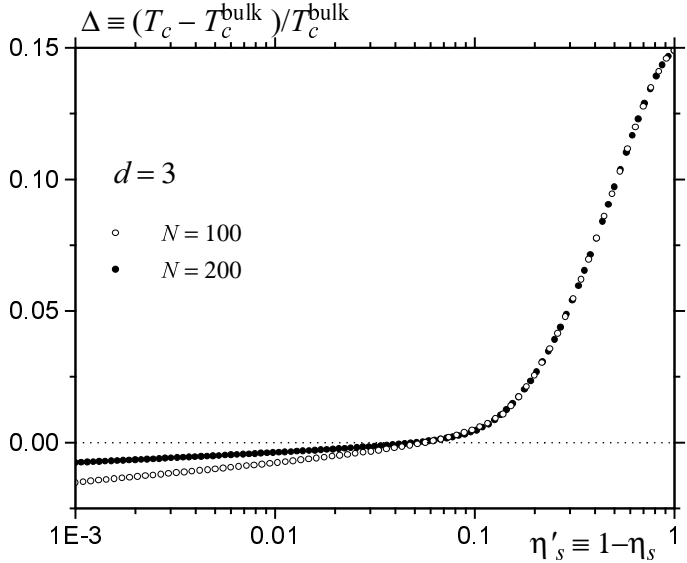,angle=-90,width=10cm}}
\end{picture}
%
\caption{ \label{s3tcet}
Curie-temperature of  ferromagnetic films with simple cubic structure  vs
surface anisotropy.
}
\end{figure}

The thickness dependence of Curie-temperature shifts in films with small surface
anisotropies are shown in Fig.\ \ref{s3tclog} in the log scale.
For $N\gg 1$ they can be represented by the formula
%
\begin{equation}\label{TcShiftSC}
\theta_c^{-1}(N) \cong \theta_{c,\rm bulk}^{-1} + \frac{ 3 }{ \pi N } \ln \frac{ 1 }{ c_3 \kappa_s },
\qquad \kappa_s \equiv \sqrt{ 2d'(1/\eta_s-1)} , 
\end{equation}
with  $\theta_{c,\rm bulk}^{-1} \equiv P_3(1)=1.51639$ and the fitting parameter $c_3\approx 1.36$.
This result, which will be derived analytically in the next section, is
simpler than that for the model with the bulk anisotropy
\cite{gar96jpal,gar98sffb}.
The latter has the form ($\kappa_c N \ll 1$)
%
\begin{equation}\label{TcShiftSCBulk}
\theta_c^{-1}(N)  \cong \theta_{c,\rm bulk}^{-1}  +  \frac{ 3 }{ \pi N } \ln\frac{ 1 }{ a_3 \kappa_c N},
\qquad  \kappa_c \equiv \sqrt{ 2d(1/\eta-1)} ,
\end{equation}
with $a_3\approx 0.35$, and $N$ under the
logarithm makes the thickness dependence of $\theta_c$ shift substantially
faster than $1/N$ (see Fig.\ \ref{s3tclog}).
For $\kappa_c N \gtrsim 1$ the bulk-anisotropy model shows a crossover to the finite-size-scaling
regime described by Eq.\ (\ref{TcShNbig}) with $\lambda=2$.
No such a crossover occurs for the model with surface anisotropy.

In Fig.\ \ref{s3tcet} the Curie temperatures of 100- and 200-layer
ferromagnetic films with simple cubic structure are shown as function of the surface anisotropy.
The film Curie temperature becomes greater than the bulk one for $\eta'_s \gtrsim
0.05$.
In this range it becomes independent of the film thickness, which is in accord with
the surface character of the phase transition.
Below the critical value of the surface anisotropy, the film Curie temperature
falls below $\theta_c^{\rm bulk}$.
One can clearly see both the log dependence of the $\theta_c$ shift on the
surface anisotropy and the $1/N$ dependence on the film thickness, as is given
by Eq.\ (\ref{TcShiftSC}).
More careful analysis shows (see Sec.\ \ref{sec_AnisLarge}) that the critical
value of the surface anisotropy, which is defined from the condition 
$T_c(N,\eta'_{s,c})=T_c^{\rm bulk}$ tends to zero with the increase of the
film thickness for $d\leq 3$.
In three dimensions this dependence is logarithmic: 
$\eta'_{s,c}(N) \sim 1/\ln N$. 
This means that Eq.\ \ref{TcShiftSC} is valid for sufficiently small anisotropy,
$\eta'_s \ll \eta'_{s,c}(N)$, or, in other words, in the thickness range 
$N\exp(-1/\eta'_s) \lesssim 1$.
For whatever small value of $\eta'_s$, it will break down for very large $N$. 
Deviation of the numerically calculated points for $\eta'_s=10^{-2}$ in Fig.\ \ref{s3tclog}
downwards from the straight line corresponding to Eq.\ (\ref{TcShiftSCBulk})
is a manifestation of this incipient breakdown.

\section{Isotropic and weakly anisotropic films}
\label{sec_AnisSmall}
\eqreset

To get an idea about ordering in films with small surface anisotropies in
$d\leq 3$ dimensions, it is convenient to start with isotropic films.
These films cannot order for any finite thickness $N$ because they are systems
of dimension $d'\leq 2$ and thus long-wavelength thermal fluctuations preclude ordering. 
On the other hand, it is physically clear that immediately below the bulk
value of the Curie temperature the susceptibility of a thick film should become
extremely high. 
This means that the lowest eigenvalue $\mu_1$ in Eq.\ (\ref{signn'mat})
becomes extremely close to zero.
The contribution of this eigenvalue dominates in the the constraint relation
(\ref{SConstrMat}), and this makes  possible analytical calculation of $\mu_1$. 
Since for the isotropic model there is no difference between longitudinal and
transverse CFs, we will use here more compact notations $E_\rho$ and $\psi_{n\rho}$
[see Eq.\ (\ref{sigzzmat})].
First, Eq.\ (\ref{SConstrMat}) can be summed over all layers with the use of
the orthogonality of wave functions $\sum_n \psi_{n \rho } \psi_{n\rho'} =\delta_{\rho\rho'}$, which
yields
%
\begin{equation}\label{SConstrSum}
 \sum_{\rho=1}^N 
\frac{ 2d' }{ 2d' +E_\rho } P_{d'}\left(\frac{ 2d' }{ 2d' + E_\rho } \right) 
= \frac{ d'N }{ d\theta }.  
\end{equation}
Next, one can subtract these equations for $\theta_c^{\rm bulk}$ and $\theta$
from each other and separate the leading term with very small $E_1(\theta)$.
This yields
%
\begin{equation}\label{P1Sep}
P_{d'}\left(\frac{ 2d' }{ 2d' + E_1(\theta) } \right) - \Sigma_N \cong  \frac{ d' N }{ d } 
\left( \frac 1\theta - \frac{ 1 }{ \theta_c^{\rm bulk} } \right) .
\end{equation}
Since $E_\rho$ with $\rho \leq 2$ are expected to change not so strongly as
$E_1$ at the temperature interval $\theta_c^{\rm bulk}-\theta$, the
quantity $\Sigma_N$ can be expected to be subdominant in comparison 
to other parts of Eq.\ (\ref{P1Sep}).

For the simple cubic lattice, $P_2$ is the Green function of the
square lattice which is given by $P_2(X) \cong (1/\pi) \ln[ 8/(1-X)]$
for $X\cong 1$.
Adopting this in Eq.\ (\ref{P1Sep}) and exponentiating yields
%
\begin{equation}\label{E1exp}
E_1(\theta) \cong C_N(\theta) \exp\left[ -\frac{ 2\pi N }{ 3 } 
\left( \frac 1\theta - \frac{ 1 }{ \theta_c^{\rm bulk} } \right) \right],
\end{equation}
where
%
\begin{equation}\label{CNDef}
C_N(\theta) = E_1(\theta_c^{\rm bulk})
\prod_{\rho=2}^N \frac{ E_\rho(\theta_c^{\rm bulk}) }{ E_\rho(\theta) } .
\end{equation}
Keeping high-lying eigenvalues with $\rho \sim N$ in the above formula is not
justified, because $P_2(X)$ does not have its asymptotic form above in this case.
On the other hand, the latter change negligibly for
$\theta$ close to $\theta_c^{\rm bulk}$ and thus the corresponding numerators
and denominators in Eq.\ (\ref{CNDef}) cancel each other.
The low-lying eigenvalues also cannot change significantly in this temperature
interval, thus the product in Eq.\ (\ref{CNDef}) should be of order unity.
This leads to the order-of-magnitude estimation 
%
\begin{equation}\label{CNEstim}
C_N \sim E_1(\theta_c^{\rm bulk}) \sim C/N^2,
\end{equation}
which is sufficient for our purposes, since $C_N$ will enter under the
logarithm in the expression for the shift of the Curie temperature of the film.
The second step in Eq.\ (\ref{CNEstim}) can be justified as follows.
For thick films at the bulk criticality, $G_n$ is close to 1 in the
main part of the film, excluding the regions near the surfaces.
Thus for estimation of the eigenvalues one can set $G_n=1$ in the whole film,
which amounts to the approximation 
$E_1(\theta_c^{\rm bulk}) \sim E_1^{\rm MFA}(\theta_c^{\rm bulk})$.
Solution of the Schr\"odinger equation (\ref{SchrEq}) with the 
potential $V_n=0$ yields eigenvalues
%
\begin{equation}\label{ErhoMFA}
E_\rho^{\rm MFA}(\theta_c^{\rm bulk})  = 
2 ( 1 - \cos q_\rho ), 
 \qquad q_\rho \equiv \pi\rho/(N+1).
\end{equation}
[so that $E_1^{\rm MFA}(\theta_c^{\rm bulk}) \sim 1/N^2$ for $N\gg 1$]
and eigenfunctions
%
\begin{equation}\label{psirhoMFA}
\psi_{n\rho} = C_{N\rho} \sin(nq_\rho), \qquad C_{N\rho} \sim 1/\sqrt{N},
\end{equation}
which describe a particle  hopping in a rigid box.

\begin{figure}[t]
\unitlength1cm
\begin{picture}(15,7.5)
\centerline{\epsfig{file=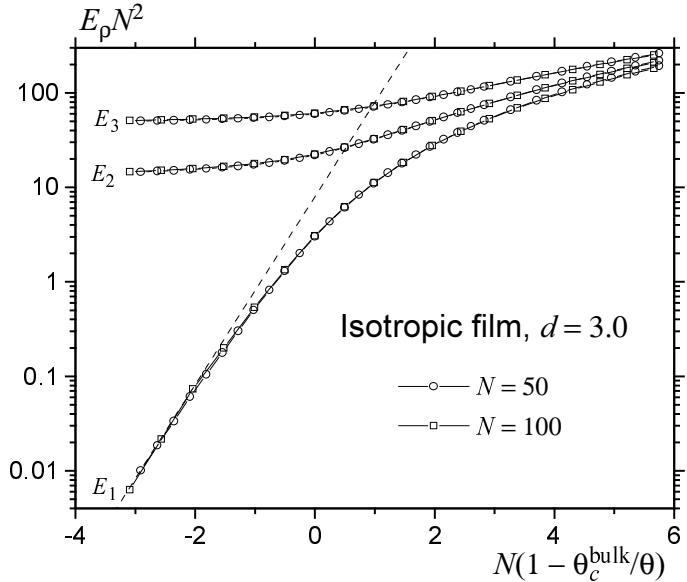,angle=-90,width=10cm}}
\end{picture}
%
\caption{ \label{s30eiso}
Temperature dependence of three lowest eigenvalues $E_\rho$ for the
isotropic film in $d=3.0$ dimensions.
Dashed lines represent Eq.\ (\protect\ref{E1exp}) with $\pi\Rightarrow 2$ and $c_{N}=8/N^2$.
}
\end{figure}

The picture described above is confirmed by numerical calculations, the results of
which are shown in Fig.\ \ref{s30eiso}.
The latter were performed for the continuous-dimension model in $d=3.0$.
The dashed line in Fig.\ \ref{s30eiso} represents Eq.\ (\ref{E1exp}), where the
transition from the sc lattice to the $d=3.0$ lattice is done by the
replacement $\pi \Rightarrow 2$, according to the general rule which can be
found in Refs.\ \cite{gar98pre,gar98sffb}. 
The constant $C$ in Eq.\ (\ref{CNEstim}) fits to 8 in $d=3.0$ dimensions.

One can ask how the variation of the gap parameter $G_n$ in the isotropic film below the
bulk criticality looks like.
The answer in the limit $\theta \ll 1 $ follows from the observation that all
 spins become strongly correlated and thus all $\sigma_{nn'}$ become nearly the same
for $q=0$.
Then from Eq.\ (\ref{CFfd}) immediately follows that $b_1=b_N \cong \onehalf$ and
$b_n \cong 1$ inside the film.
This yields
%
\begin{equation}\label{GnTZero}
G_n \cong
\left\{
\begin{array}{ll}
2d/(2d-1),                      & n=1,N                           \\
1,		               & n\neq 1,N 
\end{array}
\right. 
\end{equation}
for $\theta \ll 1$.
The corresponding zero-temperature eigenvalues can be calculated analytically
and read
%
\begin{equation}\label{ErhoTZero}
E_\rho(0)  = 2 ( 1 - \cos \bar q_\rho) , \qquad \bar q_\rho = \pi(\rho-1)/N.
\end{equation}
These eigenvalues are all shifted downwards with respect to those of
Eq.\ (\ref{ErhoMFA}), and the lowest eigenvalue is exactly zero, in accord with Eq.\ (\ref{E1exp}).
The eigenfunction of this eigenvalue is constant throughout the film:
$\psi_{n,1} = 1/\sqrt{N}$.
This is due to attraction of the particle to the potential wells at the boundaries of
the box: $V_1 = V_N = - 1$.
Note that using Eq.\ (\ref{ErhoTZero}) and Eq.\ (\ref{ErhoMFA}) in Eq.\ (\ref{CNDef})
yields $C_N =O(1)$ at $\theta \ll 1$, in contrast to estimation
(\ref{CNEstim}) just below the bulk criticality.

Calculation of the variation of the gap parameter $G_n$ in the film at $\theta < \theta_c^{\rm bulk}$
is an analytically intractable nonlinear problem, and the result of Eq.\ (\ref{E1exp})  
does not help much.
Linearization at $\theta\ll 1$ shows that deviations of $G_n$ from the
zero-temperature result of Eq.\ (\ref{GnTZero}) are linear in temperature.
A compact analytical solution can be only obtained for the trilayer.

\begin{figure}[t]
\unitlength1cm
\begin{picture}(15,7.5)
\centerline{\epsfig{file=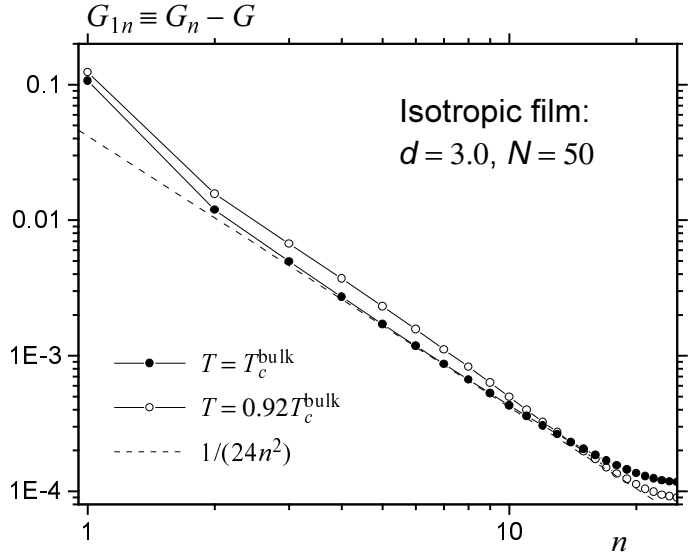,angle=-90,width=10cm}}
\end{picture}
%
\caption{ \label{sgniso}
Gap-parameter (energy-density) profile in the isotropic ferromagnetic film 
in $d=3.0$ dimensions at and below the bulk criticality.
}
\end{figure}

The deviation of the gap parameter from the bulk value, which is defined by
 $G_{1n} \equiv G_n - G$, is shown in Fig.\ \ref{sgniso} for the isotropic film in $d=3.0$
dimensions at and slightly below the bulk criticality (in both cases $G=1$).
This deviation is proportional to the nonuniform
part of the energy density \cite{gar98pre}.  
At the bulk criticality, $G_n$ has the universal form 
%
\begin{equation}\label{GnProfSemi}
G_n \cong 1 + \frac{ \onefourth -\mu^2} { 2dn^2 }, \qquad \mu =\frac{ d-3}{2},
\qquad \qquad 1 \ll n \lesssim N/4 ,
\end{equation}
for $2<d<4$, as for the semi-infinite model \cite{bramoo77prl,bramoo77jpa,gar98pre}.
This yields the large-distance form 
%
\begin{equation}\label{VnAsymp}
V_n \cong  - (\onefourth -\mu^2)/ n^2 ,
\qquad \qquad 1 \ll n \lesssim N/4 ,
\end{equation}
for potential $V_n$ in Eq.\ (\ref{SchrEq}). 
Note that in Refs.\ \cite{bramoo77prl,bramoo77jpa}, the quantity 
$V(z)\cong-(G_n-1) \cong 2dV_n$ was used.
For models with finite number of spin components, the energy-density profiles
in critical films were calculated with the help of the $\epsilon$ expansion 
\cite{eiskredie93,kreeisdie95}.
At $T=0.92 T_c^{\rm bulk}$, the profile of  $G_{1n}$ looks rather indefinite:
in the middle of the film the tendency to the zero-temperature solution of Eq.\ (\ref{GnTZero})
is already seen, whereas closer to the boundaries $G_{1n}$ still increases
with lowering temperature.
In the whole range of $n$, the relative deviation of  $G_{1n}$ from the
bulk-criticality result is of order one.
On the other hand, at such temperature the argument of the exponential in Eq.\ (\ref{E1exp}) 
is already $-10$, thus $E_1$ is very small  and futher lowering of the
temperature leads to the instability of the numerical algorithm.
Fortunately, the problem of finding the temperature variation of the gap-parameter profile in the film
below the bulk Curie temperature becomes nonessential in the physically relevant three-dimensional case,
because here the suppression of $T_c$ of the film is not so strong (see below).

As we have see above, in isotropic films in $d\leq 3$ dimensions 
 $E_1$ is very small in a wide range of temperatures but turns to zero
only at $T=0$.
If there is a small anisotropy in the system, the basic equations for the transverse
CF $\sigma_{nn'}$, Eqs.\ (\ref{CFfd}) and (\ref{sconstrfd}), are slightly modified, and the variation
of the gap parameter $G_n$ in the film slightly changes.
These changes can be found perturbatively, although it is not easy to do analytically.
When $G_n$ is inserted to the equation for the longitudinal CF
$\sigma_{nn'}^{zz}$,  perturbations of $G_n$ perturb, in turn,
$E_1$.
Since $E_1(\theta)$ goes almost horizontally, a small anisotropy is
sufficient to cause  $E_1$ to cross the zero level at a transition
temperature that is not small.

The first step, finding the perturbed variation of the gap parameter $G_n$,
can be done qualitatively in the following way.
If surface and bulk anisotropies, $\eta'_s \equiv 1-\eta_s$ and $\eta' \equiv
1-\eta$, are small, one sets ${\bf q}=0$ in the last term of Eq.\ (\ref{bnMo}),
since this term creates a gap in the transverse CF $\sigma_{nn}$ and it should
be essential at small wave vectors.
After that, defining $G_n^{(0)}$ as the solution of the isotropic problem, one
immediately finds that $G_n$ adjusts so that $\tilde b_n$ retains its isotropic value, i.e.,
%
\begin{equation}\label{GnPert}
 d/(\eta G_n)  + ( 1 -  \eta_s / \eta )d' (\delta_{n,1} + \delta_{nN} ) = d/G_n^{(0)}. 
\end{equation}
This defines the correction to $G_n$ due to anisotropy, which are positive.
Now, proceeding to the longitudinal CF, one can write for the eigenvalue
problem of Eq.\ (\ref{SchrEq}) $V_n = V_n^{(0)} + V_n^{(1)} $, where
%
\begin{equation}\label{Vn1}
V_n^{(1)} = - 2d(1-\eta)/G_n^{(0)} - 2d'(\eta-\eta_s) (\delta_{n,1} +
\delta_{nN} ) .
\end{equation}
Numerical calculations show, however, that the surface part of this
perturbation is somewhat oversimplified.
It is not strictly localized in the boundary layer but redistributed over some
region, decaying in three dimensions slightly faster than $1/n^3$, presumably as
$1/(n^3 \ln n)$.
This feature in nonessential for the determination of the $T_c$ shift below;
the difference of the result with respect to those obtained with the use of
the simplified form of Eq.\ (\ref{Vn1}) will be absorbed into analytically
unknown numerical factors. 

\begin{figure}[t]
\unitlength1cm
\begin{picture}(15,7.5)
\centerline{\epsfig{file=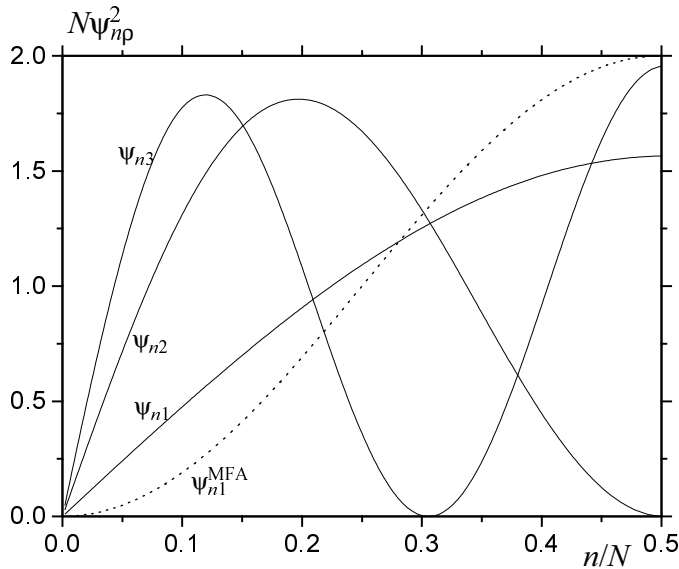,angle=-90,width=10cm}}
\end{picture}
%
\caption{ \label{s30wf0}
Numerically calculated squares of normalized eigenvectors $\psi_{n\rho}$ for
the isotropic film ($N= 500$) in $d=3.0$ dimensions at bulk criticality.
Linear behavior at small $n$ is in accord with Eq.\ (\protect\ref{psirhoBess}).
The mean-field result of Eq.\ (\protect\ref{psirhoMFA}) is shown by the dotted line.
}
\end{figure}

The first-order corrections to $E_\rho$  
due to the diagonal perturbation $V_n^{(1)}$ have the form
%
\begin{equation}\label{Erho1}
E_\rho^{(1)} = \sum_{n=1}^N V_n^{(1)} \psi_{n\rho}^2 ,
\end{equation}
as in the usual quantum-mechanical perturbation theory.
In the temperature range of interest, slightly below the bulk criticality, the
variation of the gap parameter $G_n$ does not strongly differ from that at
bulk criticality.
Thus estimation of $\psi_{n\rho}^2$ in Eq.\ (\ref{Erho1}) can be done at
$T=T_c^{\rm bulk}$.
Here not too close to the surfaces and to the middle of the film one should
consider the Schr\"odinger equation  (\ref{SchrEq}) with the
potential $V_n$ given by Eq.\ (\ref{VnAsymp}).
The standing-wave solution of this equation for $n\gtrsim 1$ in the semi-infinite
geometry can be expressed through the Bessel functions and labeled with the
 wave vector taking continuous values from the interval $(0,\infty)$ 
\cite{bramoo77prl,bramoo77jpa,gar98pre}.
In the film this wave vector becomes quantized, and the normalized expression
for $\psi_{n\rho}$ in the region $n \lesssim N/2$ reads
%
\begin{equation}\label{psirhoBess}
\psi_{n\rho} \sim (nq_\rho/N)^{1/2} J_\mu(nq_\rho), \qquad q_\rho \sim \pi\rho/N.
\end{equation}
Thus for using $J_\mu(z) \sim z^\mu$, $z\ll 1$,  one obtains 
%
\begin{equation}\label{ProbabBess}
\psi_{n,1}^2 \sim n^{1+2\mu}/N^{2(1+\mu)}, \qquad \psi_{1,1}^2 \sim 1/N^{2(1+\mu)}.
\end{equation}
For comparison, the MFA solution of Eq.\ (\ref{psirhoMFA})  yields $\psi_{11}^2
\sim 1/N^3$.
Strong increase of the probability of finding the particle near the boundaries
for $N\gg 1$ in our case, which is illustrated in Fig.\ \ref{s30wf0}, is due to
the long-range attractive potential $V_n$.
Note that at low temperatures, where $G_n$ approaches its limiting form 
given by Eq.\ (\ref{GnTZero}), this effect becomes even stronger:  $\psi_{n,1}^2 \cong 1/N$.
But there are no bound states near the surface in the isotropic model at any
temperatures.

Now, from Eq.\ (\ref{Erho1}) one obtains for the surface- and bulk-anisotropy
models
%
\begin{equation}\label{Erho1SuBu}
E_1^{(1, {\rm surface})} \sim -\kappa_s^2/N^{2(1+\mu)},
\qquad E_1^{(1, {\rm bulk})}  \cong -\kappa_c^2,
\end{equation}
where in the bulk case $G_n^{(0)} \cong 1$ in the main part of the film and the normalization of eigenvectors 
$\psi_{n\rho}$ has been used.
The Curie temperature of the film can be found from Eq.\ (\ref{TcEqMat}) in the
form $E_1^{(0)} + E_1^{(1)} =0$, where $E_1^{(0)}$  is given  for $d=3$ by Eqs.\ (\ref{E1exp})
and  (\ref{CNDef}).
Explicitly, one has 
%
\begin{equation}\label{TcEqPert}
\left\{ \kappa_s^2 \atop (\kappa_cN)^2 \right\} 
\sim  \exp\left[ -\frac{ 2\pi N }{ 3 } 
\left( \frac 1\theta_c - \frac{ 1 }{ \theta_c^{\rm bulk} } \right) \right]
\end{equation}
for the surface- and bulk-anisotropy models, respectively.
This results in Eqs.\ (\ref{TcShiftSC}) and  (\ref{TcShiftSCBulk}), where
the numbers $c_3$ and $a_3$ cannot be found analytically and should be fitted
to the numerical solution.
Remember that this analytical scheme for determination of $\theta_c$ works if
the argument of the exponential above is large.
The method evidently breaks down for the bulk-anisotropy model, if $\kappa_c N
\gtrsim 1$.
Here the result for $\theta_c$  crosses  over to the finite-size-scaling solution of
Eq.\ (\ref{TcShNbig}) with $\lambda=2$ and $A\sim 1/\kappa_c$ \cite{gar96jpal,gar98sffb}. 
For the model with surface anisotropy, Eq.\ (\ref{TcEqPert}) also breaks down
at sufficiently large $N$ due to the surface phase transition,  which will be
considered in more detail in the next section.

For very small anisotropy, the film orders at the temperature $\theta_c \ll \theta_c^{\rm bulk} \sim 1$, 
where $\psi_{n,1}^2 \cong 1/N$ and in Eq.\ (\ref{E1exp}) $C_N = O(1)$ [see comment 
after Eq.\ (\ref{GnTZero})].
This yields 
%
\begin{equation}\label{TcShiftSCLT}
\theta_c^{-1}(N) \cong  \frac{ 3 }{ \pi N } \ln \frac{ \sqrt{N} }{ \kappa_s } ,
\qquad \theta_c^{-1}(N)  \cong  \frac{ 3 }{ \pi N } \ln\frac{ 1 }{ \kappa_c }
\end{equation}
for the surface- and bulk-anisotropy models, respectively, in accord with Eqs.\ (\ref{TcAnSurfLo}) and (\ref{TcAnLo}).
It should be stressed that the applicability conditions for the formulae above are difficult to fulfill for thick
films, $N \gg 1$.
For the latter, the shift of the Curie temperature is typically small and  Eqs.\ (\ref{TcShiftSC}) and 
(\ref{TcShiftSCBulk}) are relevant.

Let us consider now ferromagnetic films in dimensions lower than three.
For the continuous-dimension model \cite{gar98pre} the lattice Green function of layers, $P_{d'}$,
is given by 
%
\begin{equation}\label{Pd'Def}
P_{d'}(X) = \frac{ d' }{ \Lambda^{d'} } \int_0^\Lambda 
\frac{ q^{d'-1} dq }{ 1 - X\lambda'_q } , \qquad \lambda'_q \cong 1 - q^2/(2d').
\end{equation}
For $X$ close to 1 this yields
%
\begin{equation}\label{Pd'Res}
P_{d'}(X) \cong
\left\{
\begin{array}{ll}
C_{d'}/\kappa_{d'}^{2-d'},                      & d'<2                           \\
W_{d'} + C_{d'}\kappa_{d'}^{d'-2},	 & d'>2,
\end{array}
\right.
\qquad \kappa_{d'} \equiv \sqrt{2d'(1/X -1)} \ll 1 ,
\end{equation}
where the Watson integral $W_{d'}$ and the coefficient $C_{d'}$ are given by
%
\begin{equation}\label{Wd'Cd'Def}
W_{d'} = \frac{ (d')^2 }{ (d')^2 - 4 } , 
\qquad C_{d'} = \frac{ d' }{ \Lambda^{d'} } \frac{ \pi  d' }{ \sin[(2-d')\pi/2] } .
\end{equation}
In three dimensions, the exact result is $P_{3.0'} = [1/(2X)] \ln [(1+X)/(1-X)]$.
One should not mix up $P_{3.0'}$ (two continuous dimensions) with $P_{2.0}$ (one discrete dimension 
and one continuous dimension), etc.
 
For $d<3$ the first term of Eq.\ (\ref{P1Sep}) is of order $1/E_1^{(3-d)/2}$
and it dominates over $\Sigma_N$.
The latter is determined by other low-lying eigenvalues which are of order $E_\rho\sim (\rho/N)^2$.
Thus $\Sigma_N \sim N^{3-d}C_{d'}$. 
This correction  term will be retained in the formulae in order to provide correct limiting transition $d\to 3$.
Using Eq.\ (\ref{Pd'Res}) and equating $E_1$ to the anisotropic correction $E_1^{(1)}$ with 
the opposite sign, one obtains
%
\begin{equation}\label{Tcdlo0}
\theta_c^{-1} \cong  \theta_{c,\rm bulk}^{-1}  + \frac{ d C_{d'} }{ d' N } 
\left( \frac{ 1 }{ (-E_1^{(1)})^{(3-d)/2 }} - \frac{\Sigma_N }{C_{d'} } \right).
\end{equation}
Inserting here expressions for $E_1^{(1)}$ from Eq.\ (\ref{Erho1SuBu}) and using the value of $\mu$
from Eq.\ (\ref{GnProfSemi}), one arrives at the final results
%
\begin{equation}\label{TcdloSu}
\theta_c^{-1} \cong  \theta_{c,\rm bulk}^{-1}  +  
\frac{ d' }{ \Lambda^{d'} } \frac{ \pi  d }{ \sin[(3-d)\pi/2] }
\frac{  1 - (c_d \kappa_s)^{ 3-d } N^{(3-d)^2/2}  }
{  (\bar c_d \kappa_s)^{ 3-d } N^{d-2 +(3-d)^2/2}  } 
\end{equation}
for the surface-anisotropy model and 
%
\begin{equation}\label{TcdloBu}
\theta_c^{-1} \cong  \theta_{c,\rm bulk}^{-1}  + 
 \frac{ d' }{ \Lambda^{d'} } \frac{ \pi  d }{ \sin[(3-d)\pi/2] }
\frac{  1 - ( a_d \kappa_c N)^{3-d}  }
{  \kappa_c^{ 3-d } N  }
\end{equation}
for the bulk-anisotropy model.
Here $c_d$, $\bar c_d$, and $a_d$ are numbers that should be fitted to the numerical solution. 
One can check that for $d\to 3$ the formulae above go over to Eqs.\ (\ref{TcShiftSC}) and 
(\ref{TcShiftSCBulk}) (the additional factors $2/\pi$ in the latter are due to the differense between 
$d=3.0$ and $d=3$ models).
Moreover, both Eqs.\ (\ref{TcdloSu}) and  (\ref{TcdloBu}) cross over to the single result
$\theta_c^{-1} - \theta_{c,\rm bulk}^{-1} \sim 1/N^{d-2}$
\cite{o'cstebra97,gar98sffb} in dimensions above 3, which is well-defined in the isotropic limit.

\begin{figure}[t]
\unitlength1cm
\begin{picture}(15,7.5)
\centerline{\epsfig{file=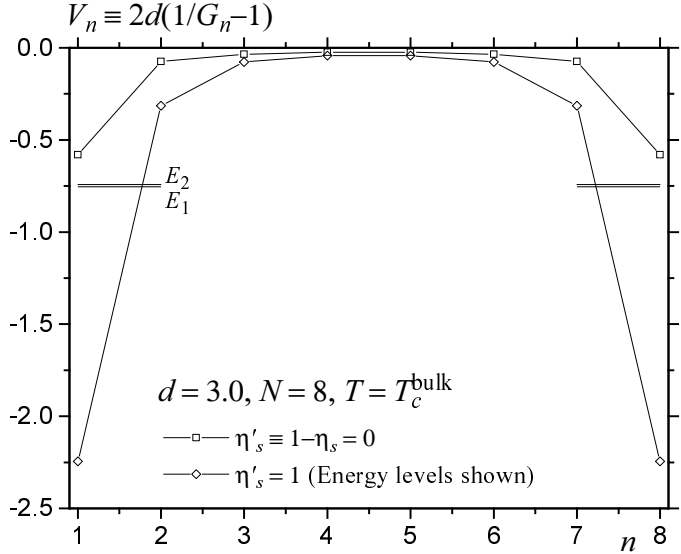,angle=-90,width=10cm}}
\end{picture}
%
\caption{ \label{swe8en}
Numerically calculated effective potentials $V_n$ for the isotropic film ($\eta'_s=0$) and that with
the extreme surface anisotropy, $\eta'_s=1$, for $N=8$ in $d=3.0$ dimensions
at the bulk criticality.
}
\end{figure}
\begin{figure}[t]
\unitlength1cm
\begin{picture}(15,7.5)
\centerline{\epsfig{file=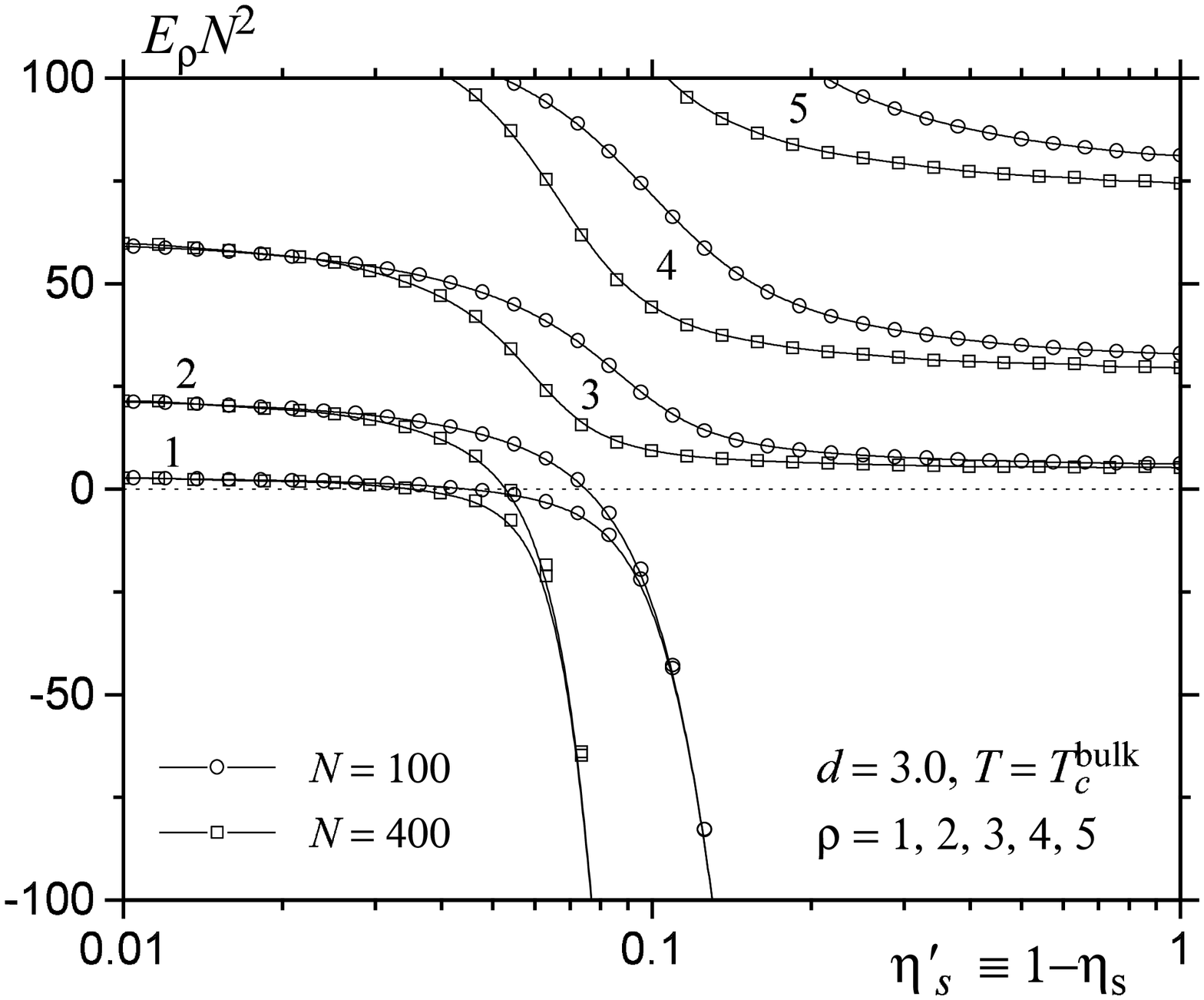,angle=-90,width=10cm}}
\end{picture}
%
\caption{ \label{s30lam}
Surface-anisotropy dependence of the lowest energy levels of the potential $V_n$ 
for $N=100$ and $N=400$ films at the bulk criticality.
}
\end{figure}

\section{Surface phase transition in films}
\label{sec_AnisLarge}
\eqreset

The surface of a semi-infinite magnetic system orders at a temperature above
the bulk Curie temperature if there is a bound surface state of the Schr\"odinger
equation (\ref{SchrEq}), which lies below the continuum of the delocalized
(bulk) states, i.e., $E_1<V_\infty$.
In this case, with lowering temperature $E_1$ reaches the zero value before
all other (bulk)  eigenvalues, and it dominates the longitudinal
susceptibility $\chi_{zn}= \sigma_{nn}^{zz}({\bf q}=0)/\theta$  [see Eq.\ (\ref{sigzzmat})] 
in the boundary region, where the eigenfunction $\psi_{n1}$ is localized.
An example of the surface bound states is shown in Fig.\ \ref{swe8en} for the
$N=8$ film in $d=3.0$ dimensions at the bulk criticality.
For the isotropic model, the potential wells near the surfaces are not strong
enough to create a bound state.
In contrast, for the extremely strong surface anisotropy the wells are deeper
and there are bound
states in each of the wells, which show a small tunnel splitting.
Both models possess bulk states with $E_\rho >0$, which are not shown.
The surface-anisotropy dependence of several lowest energy levels for thick
films at the bulk criticality is shown in Fig.\ \ref{s30lam}.
One can see that the energy levels $E_\rho >0$ nearly scale with $1/N^2$, which is 
characteristic for the bulk.
Negative energies correspond to the states localized near the surfaces, here
$E_{1,2}$ are nearly degenerate and practically independent of $N$. 

A natural question is how strong the surface anisotropy should be to create a
potential well which can accomodate a bound state.
The well known result in quantum mechanics (see, e.g., Ref.\ \cite{lanlif3})  is that in one dimension a whatever
small potential well creates a bound state with the energy quadratic in the
volume of the well: $E \cong -\onefourth \left[\int_{-\infty}^\infty V(x) dx \right]^2$
for $\hbar=1$ and $m=1/2$.
If, however, the potential well is situated near a potential hump or a near an
rigid wall, localization of the particle costs additional potential or
kinetic energy, respectively, and it requires that the well strength exceeds
some critical value.
In this case  the result is 
%
\begin{equation}\label{EBoundGen}
E\cong -A(P-P_c)^\zeta + \rm const,
\end{equation}
where $P$ is appropriately determined strength of the well and for {\em short-range} potential wells $\zeta=2$.
In the particular case of a rectangular well of depth $V_0$ and width $a$,
which is sided by a rigid wall, one has $P=a\tilde q$, $\tilde q \equiv \sqrt{V_0}$, $P_c = \pi/2$, and
$A=\pi^2/(2a)^2$.
If the potential $V(x)$ has a long tail, the situation becomes more
complicated, and the exponent $\zeta$ deviates from 2, as we shall see below.

Calculation of the critical strength $P_c$ in Eq.\ (\ref{EBoundGen}) for our
problem (\ref{SchrEq}) requires, in general,  knowing the potential $V_n$ in
the whole range of $n$ including the surface region, $n \sim 1$.
Whereas at the bulk criticality the asymptotic form of $V_n$ is given by
Eq.\ (\ref{VnAsymp}), the values of $V_n$ for $n\sim 1$ can be only determined
numerically \cite{gar98pre}.
Nevertheless, it can be shown that the isotropic semi-infinite model at the
bulk criticality in $d\leq 3$ 
dimensions is in the critical state, $P=P_c$.
A whatever small surface anisotropy $\eta'_s$ makes the well deeper in the
region $n\sim 1$ and it thus creates a surface bound state.
As was argued above, this leads to the surface phase transition.
This strong result follows from the form of the spin CF, which in the
asymptotic region $n,n'\gg1 $ for $q\ll 1$ reads \cite{bramoo77prl,bramoo77jpa,gar98pre}
%
\begin{equation}\label{sigik1}
\sigma_{nn'}({\bf q}) =  2d\theta
\left\{ 
\begin{array}{ll}
\displaystyle
\sqrt{nn'} I_\mu(qn) K_\mu(qn'),       	& n\leq n'                           \\
\displaystyle
\sqrt{nn'} I_\mu(qn') K_\mu(qn),                      & n'\leq n ,                       \\
\end{array}
\right. 
\end{equation}
with $\mu$ defined by Eq.\ (\ref{GnProfSemi}).
Far from the boundary, $qn, qn' \gg 1$,  this CF reduces to its bulk value, 
$\sigma_{nn'}({\bf q}) = (d\theta/q) \exp(-q|n-n'|)$.
In the region $n,n'\sim 1$ Eq.\ (\ref{sigik1}) is modified by nonsingular
factors of order unity\cite{gar98pre}.
The spin CF above is proportional to the Green function which can be used to
calculate perturbations of the solution of the problem (\ref{SchrEq}), with $E=q^2$, due to
small perturbations of the potential $V_n$.
Such a perturbation theory, fails, however, in the bulk, since the bulk Green
function above also diverges for $q\to 0$. 
A whatever small perturbation of $V_n$ changes the wave functions with $E\to 0$
in a nonperturbative way, which leads to formation of bound states for
attractive perturbations \cite{lanlif3}.
To analyze the semi-infinite problem, one can use
%
\begin{eqnarray}\label{ikzsmall}
&&
I_\mu(z) \cong  \frac{ 1 }{ \Gamma(1+\mu) } \left( \frac z2 \right)^\mu [1+O(z^2)],
\qquad z \ll 1
\nonumber\\
&&
K_\mu(z) = \frac{ \pi }{ 2\sin(\pi\mu) } [ I_{-\mu}(z) - I_\mu(z) ]
\end{eqnarray}
for the modified Bessel and Macdonald functions.
One can see that for $\mu \leq 0$ (i.e., $d\leq 3$) the Green function above
diverges in the limit $q\to 0$ (for $d=3$ logarithmically), whereas for $\mu > 0$ (i.e., $d > 3$) it
remains finite in this limit.
Thus, in $d>3$ dimensions there should be a critical value of the surface
anisotropy, $\eta_{s,c}$, above which there is a surface phase transition,
whereas for $d\leq 3$ one has $\eta_{s,c}=0$.

Different behavior for $d>3$ and $d<3$ observed above is entirely due to
the different forms of $V_n$ for $n\sim 1$, whereas in the asymptotic region
$n\gg 1$ the potential $V_n$ given by Eq.\ (\ref{VnAsymp}) is the same below
and above three dimensions.
If one goes away from $d=3$ in both directions, the attractive tail of $V_n$ weakens, but 
for $d<3$ the depth of the well increases in the surface region, $n\sim 1$, 
(see Fig.\ 1($b$) of Ref.\ \cite{gar98pre}), so
that the well always remains in the critical state.
In the limit $d\to 2$ the attractive tail of $V_n$ disappears, and the
variation of the gap parameter $G_n$ approaches Eq.\ (\ref{GnTZero}), which
corresponds to $V_1=V_N = -1$, $V_n=0$ ($n\neq 1,N$).
It can be checked directly that a whatever small further decrease of the boundary
value of this potential leads, for the semi-infinite problem, $N=\infty$, to the formation of a bound state.
Determination of $V_n$ for $n\sim 1$ is an analytically intractable nonlinear
problem.
Nevertheless, Bray and Moore \cite{bramoo77prl,bramoo77jpa} could obtain the
spin CF of Eq.\ (\ref{sigik1}), which has {\em different} forms for $d>3$ and
$d<3$ and contains the relevant information, without explicitly analyzing the
region $n \sim 1$!

Now let us analyze how the energy of the surface bound state depends on the strength
of the potential well if the latter slightly exceeds its critical value.
For simplification, we will consider, instead of Eq.\ (\ref{SchrEq}), a continuous Schr\"odinger equation
$-\psi'' +V(x) \psi = E \psi$ with the potential $V(x)$ modelled as
%
\begin{equation}\label{VxModel}
V(x) =  
\left\{ 
\begin{array}{ll}
\infty,    	& x< 0 \\
-V_0,       	& 0 \leq x \leq a                          \\
-(\onefourth - \mu^2)/x^2,                      & x > a                        \\
\end{array}
\right. 
\end{equation}
[cf Eq.\ (\ref{VnAsymp})].
If we choose $a=\pi/2$,  then for $d=2$ the long tail of $V(x)$
disappears and $V_0=1$ becomes the critical depth of the potential well, as for the
original discrete problem.
In general, this method tells nothing about the critical value of the surface
anisotropy, but allows the determination of the exponent $\zeta$ in
Eq.\ (\ref{EBoundGen}).
The bound solution of the problem above, if it exists, has the form
%
\begin{equation}\label{psixModel}
\psi(x) =  
\left\{ 
\begin{array}{lll}
C_1\sin(\tilde q x),       	& 0 \leq x \leq a   &   (\tilde q \equiv \sqrt{E+V_0})                       \\
C_2 \sqrt{\tilde\kappa x} K_\mu(\tilde\kappa x),       & x > a  &
(\tilde\kappa \equiv \sqrt{-E} ) .                    \\
\end{array}
\right. 
\end{equation}
Here for very small $|E|$ one can neglect $E$ in $\tilde q$ and use the
small-argument form of $K_\mu(z)$, which follows from Eq.\ (\ref{ikzsmall}).
Then the boundary conditions at $x=a$ result in the equation determining
$\tilde\kappa$:
%
\begin{equation}\label{tildekappaEq}
\tilde q a \cot \tilde q a = \frac 12 - |\mu|  - 
\frac{ 2|\mu|  (\tilde\kappa a r_\mu)^{2|\mu|} }{ 1 -  
(\tilde\kappa a r_\mu)^{2|\mu|} },
\qquad r_\mu \equiv \frac 12 \left[ \frac{ \Gamma(1-\mu) }
{  \Gamma(1+\mu) }\right]^{1/(2|\mu|)}.
\end{equation}
Setting $\tilde\kappa=0$ determines the critical value of the well strength
$P_c$, say, its depth $V_0$.
For $P$ slightly above $P_c$, Eq.\ (\ref{tildekappaEq}) can be represented in
the form
%
\begin{equation}\label{tildekappaEq1}
B(P-P_c) \cong 
\frac{ |\mu|  (\tilde\kappa a r_\mu)^{2|\mu|} }{ 1 -  
(\tilde\kappa a r_\mu)^{2|\mu|} },
\end{equation}
which yields 
%
\begin{equation}\label{EBoundmu}
E = - \tilde\kappa^2 \cong - \frac{ 1 }{ (ar_\mu)^2 } 
\left[ \frac{ B(P-P_c) }{ |\mu| + B(P-P_c) } \right]^{1/|\mu|},
\qquad \mu \equiv \frac{ d-3 }{ 2} 
\end{equation}
for the energy of the bound state.
One can see that the ``classical'' one-dimensional behavior with the quadratic
dependence of $|E|$ on $P-P_c$ is only realized for $d=2$ and $d=4$ where 
$|\mu|=1/2$ and long tail of $V(x)$ in Eq.\ (\ref{VxModel}) disappears.
For $d=3$ Eq.\ (\ref{EBoundmu}) regularizes to the expression
%
\begin{equation}\label{EBoundd3}
E \cong  - \left( \frac{ 2 }{ a e^\gamma} \right)^2 
\exp\left[ - \frac{ 1 }{ B(P-P_c) } \right] , \qquad \gamma = 0.5772,
\end{equation}
which resembles the well known result for the energy of the bound state in two
dimensions \cite{lanlif3}.
Indeed, in two dimensions the radial part $\psi(r)$ of the wave function 
$\Psi(r,\phi)= r^{-1/2}\psi(r) \exp(\pm i m\phi)$, $m=0, 1, 2, \ldots$ for the problem {\em without} potential
energy satisfies the one-dimensional Schr\"odinger equation with the effective potential
energy written in Eq.\ (\ref{VxModel}) for $x>a$, with $\mu \Rightarrow m$
(see, e.g., Ref.\ \cite{donhouma98}). 
Now, returning to the original problem with the surface anisotropy, one can
notice that the depth of the potential wells near the surfaces change linearly
with $\eta'_s$, thus one can replace  in Eqs.\ (\ref{EBoundmu}) and 
(\ref{EBoundd3}) $P-P_c$ by $\eta'_s -\eta'_{s,c}$, where $\eta'_{s,c}=0$ for
$d\leq 3$.

The temperature of the surface phase transition, $\theta_c$,  can  now be
determined using the results above.
At  $\theta_c$, which is slightly above $\theta_c^{\rm bulk}$,  the energy of
the bound state equals zero, but the bulk level of the potential, $V_\infty$, slightly
exceeds zero.
The value of $\theta_c$ can be found equating $|E|$, which is given by
Eq.\ (\ref{EBoundmu}), to $V_\infty$:
%
\begin{equation}\label{SthetacEq}
|E| = V_\infty \sim 1-G \sim \kappa^2 \sim (\theta_c - \theta_c^{\rm bulk})^{2\nu_b},
\end{equation}
where $\nu_b = 1/(d-2)$ is the critical index for the bulk correlation length
for the $D=\infty$ model.
This yields
%
\begin{equation}\label{Sthetac}
\theta_c - \theta_c^{\rm bulk} \cong 
\left[ \frac{ B(\eta'_s -\eta'_{s,c}) }{ |\mu| + B(\eta'_s -\eta'_{s,c}) }
\right]^{1/\dot \Phi},
\qquad \dot \Phi = \frac{ |d-3| }{ d-2} .
\end{equation}
The critical index $\dot \Phi$ was calculated in Ref.\ \cite{dieeis84} for the
model with arbitrary number of spin components $n$ in the second order in 
$\varepsilon = 4-d$.
In the limit $n\to \infty$ the result of Ref.\ \cite{dieeis84} becomes 
$\dot \Phi = 1/2 -\varepsilon/4 - \varepsilon^2/8 + O(\varepsilon^3)$, which
is in accord with Eq.\ (\ref{Sthetac}).
Note, however, that the $\varepsilon$ expansion fails below three dimensions
 for the model with infinite number of spin components, which is
considered here.

\begin{figure}[t]
\unitlength1cm
\begin{picture}(15,7.5)
\centerline{\epsfig{file=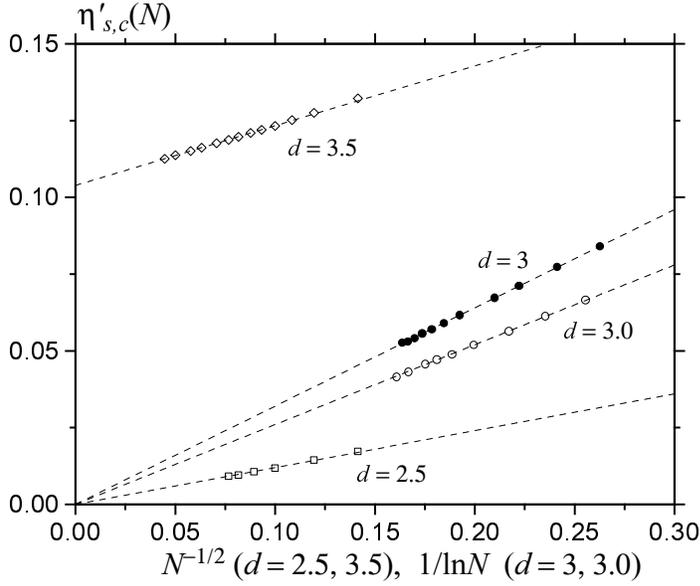,angle=-90,width=10cm}}
\end{picture}
%
\caption{ \label{setscn}
Critical surface anisotropy vs film thickness 
($50\leq N \leq 500$ for $d=3$, 3.0, 3.5 and $50\leq N \leq 170$ for $d=2.5$) 
in different dimensions.
The data correspond to the film with surface anisotropy on one of the two
surfaces.
The straight dashed lines are fits to the numerical data. 
}
\end{figure}

In films, surface bound states cannot be rigorously separated from the bulk
ones.
If these bound states are very shallow, which is the case near the special
transition ($\theta_c^{\rm surface} =\theta_c^{\rm bulk}$), the localization
length of the bound states is very large and it exceeds the thickness of the film.
Because if this finite-size effect, the critical value of the parameter which
controls the surface phase transition (here the surface anisotropy) cannot be
determined unambigouosly.
A natural choice is to define $\eta_{s,c}(N)$ from the condition
$\theta_c(N,\eta'_{s,c})=\theta_c^{\rm bulk}$.
The value of $\eta_{s,c}(N)$ can be found as the point of the intersection of
the lowest eigenvalue $E_1(N,\eta'_s)$ with the zero level at the bulk
criticality (see Fig.\ \ref{s30lam}).
For the model with symmetric surfaces, however, the second eigenvalue,  
$E_2$, also goes down, crosses the zero level at somewhat larger
value of $\eta'_s$ and then very fast becomes almost degenerate with $E_1$. 
The latter situation corresponds to the two bound states well localized on both
surfaces, with a small tunnel splitting.
Thus, crossing of $E_2(N,\eta'_s)$ with the zero level, as well as the
degeneracy of $E_1$ with $E_2$, could also be used as a criterion for the
formation of bound states and thus for the special transition.
One more and probably better possibility is to consider the film with a
surface anisotropy on only one of the two surfaces.
Here there is no complication arising from the tunneling between the bound
states across the film; only the lowest eigenvalue $E_1$ goes over to the
bound state, whereas $E_2$ remains always positive. 

In films there is no singularity in the dependence of $E_1$ on the surface
anisotropy, this dependence is linear near $E_1=0$.
On the other hand, the singularity of $E_1(\eta'_s)$ studied above for the
semi-infinite problem above mirrors in the dependence $\eta'_{s,c}(N)$.
This dependence can be obtained if one uses a potential of the type of 
Eq.\ (\ref{VxModel}) for a film, sets $\tilde \kappa=0$ and imposes the
symmetry condition on the wave function in the middle of the film, $x=L/2$.
This yields Eq.\ (\ref{tildekappaEq1}) with $P\Rightarrow P_c(L)$,
$P_c\Rightarrow P_c(\infty)$ and $\tilde \kappa \Rightarrow 2/L$.
In terms of the original variables, dropping numerical factors, one can write
%
\begin{equation}\label{etascN}
\eta'_{s,c}(N) - \eta'_{s,c}(\infty) \sim 
\frac{2 |\mu| N^{-2|\mu|} }{1-N^{-2|\mu|} } 
\;\;\Rightarrow\;\; \frac{ 1 }{ \ln N } \qquad (d=3) .
\end{equation}
This result, as well as the conjecture $\eta'_{s,c}(\infty)=0$ for $d\leq 3$
made at the beginning of this section, are confirmed by numerical calculations
the results of which are shown in Fig.\ \ref{setscn}.

Positive values of $\eta'_{s,c}(N)$, even for $d\leq 3$, reflect the general tendency
of the film to order at a temperature below the bulk Curie temperature.
The latter is the case considered in the preceding section, and now it is
clear that the applicability criterion for Eqs.\ (\ref{TcShiftSC}) and 
(\ref{TcdloSu}) is $\eta'_s \ll \eta'_{s,c}(N)$.
For $d\leq 3$, Eqs.\ (\ref{TcShiftSC}) and  (\ref{TcdloSu}) break down for
whatever small surface anisotropy $\eta'_s$, if the film thickness $N$ is
large enough.
One can see that in three dimensions $\eta'_{s,c}(N)$ decreases logarithmically slowly,
thus  Eq.\ (\ref{TcShiftSC}) works in a wide range $N \lesssim
\exp(1/\eta'_s)$ for small surface anisotropies.
The applicability range of Eq.\ (\ref{TcdloSu}) shrinks fast with the descease
of the spatial dimension $d$.

\section{Discussion}
\label{sec_Disc}
\eqreset 

In this paper it has been shown that the finite-size-scaling formula for the
$T_c$ shift in magnetic films, Eq.\ (\ref{TcShNbig}), which seems to be the
only theoretical tool for interpretation of experiments 
(see, e.g., Ref.\ \cite{libab92}), describes in fact only one of several
regimes.
For the model with bulk anisotropy, the situation depends on ratio of the film
thickness $N$ and the transverse correlation length $\xi_{c\perp}$, which is
usually ignored as a noncritical variable.
For $N\kappa_c \gtrsim 1$, where $\kappa_c \equiv 1/\xi_{c\perp}$ at
criticality, a different regime described by Eq.\ (\ref{TcShiftSCBulk}) 
\cite{gar96jpal,gar98sffb} is realized instead of Eq.\ (\ref{TcShNbig}).
For the model with surface anisotropy, which is present in many experimentally
investigated films, Eq.\ (\ref{TcShNbig}) never appears.
Instead, the $T_c$ shift follows Eq.\ (\ref{TcShiftSC}) in three dimensions for 
the surface anisotropy small enough.
If surface anisotropy exceeds the critical value, $\eta'_s > \eta'_{s,c}(N)$,
Eq.\ (\ref{TcShiftSC}) breaks down and the film orders via the surface phase
transition above the bulk Curie temperature (see Fig.\ \ref{s3tclin}).
A remarkable result is that $\eta'_{s,c}(N)$ goes to zero in the semi-infinite
limit, $N\to\infty$, for $d\leq 3$ (see Fig.\ \ref{setscn}).
That is, a whatever small surface anisotropy leads to the surface phase
transition in the bulk-isotropic semi-infinite model.
This contrasts the isotropic model with enhanced surface exchange, which does
not show any surface phase transition for $d\leq 3$.
In three dimensions, $\eta'_{s,c}(N) \sim 1/\ln N$, thus Eq.\ (\ref{TcShiftSC})
is valid in a wide range of the film thicknesses: $N\lesssim \exp(1/\eta'_s)$
for $\eta'_s \ll 1$.

One can question whether all these effects, which have been demonstrated above
for the $D=\infty$ model, survive for the realistic classical Heisenberg model,
$D=3$.
I expect that, in general, these effects should survive, because
they are due to the nearly Goldstone modes in a weakly anisotropic magnetic
system, and these Goldstone modes are inherent in all model with $D\geq 2$.
On the other hand, the nonlinear coupling of fluctuations, which arises for
finite number of spin components $D$, suppresses fluctuations to some extent.   
This can be already seen from the fact that in the bulk 
$\theta_c \equiv T_c/T_c^{\rm MFA}$ monotonically decreases with $D$ and
reaches its minimum in the spherical limit $D=\infty$ (strongest fluctuations).  
For the semi-infinite problem, the surface susceptibility $\chi_{11}$ at the
ordinary phase transition diverges for $d\leq 3$ 
(i.e., $\gamma_{11}^{\rm ord} > 0 $ for $d<3$), if $D=\infty$.
For finite $D$ the second-order $\varepsilon$ expansion 
(see, e.g., Ref.\ \cite{dieeis84} and references therein, or, for a review, 
Ref.\ \cite{die86a}) suggests that $\gamma_{11}^{\rm ord}$ remains positive at $d=3$
(no divergence of the surface susceptibility) and probably changes sign at some critical dimension lower than 3.
Thus, fluctuations are somewhat suppressed, and the situation is a bit closer to the
mean-field one ($d=4$), in comparison to the limit  $D=\infty$.
This is an indication that in three dimensions a {\em finite} value of the surface
anisotropy may be needed for the surface phase transition, in contrast to the zero
value obtained in Sec.\ \ref{sec_AnisLarge}
Computing this critical value of the surface anisotropy with the help of MC
simulations or other methods, as well as search for the regimes for the $T_c$
shift in films established above  (or rather for  their analogues for the Heisenberg
model),  seems to be an interesting problem.


\end{document}